\documentclass[twocolumn,trackchanges,floatfix]{aastex631}
\shorttitle{UV-bright BSSs in NGC\,362}
\graphicspath{{./}{figures/}}

\begin{document}
\title{ GlobULeS-IV. UVIT/{\it AstroSat} detection of  extremely low mass white dwarf companions to blue straggler stars in NGC\,362 }

\author{Arvind K. Dattatrey}
\affiliation{ Aryabhatta Research Institute of Observational Sciences,
Manora Peak, Nainital 263002, India.}
\affiliation{ Deen Dayal Upadhyay Gorakhpur University, Gorakhpur, Uttar Pradesh 273009,India.} 

\author{R.K.S. Yadav}  
\affiliation{ Aryabhatta Research Institute of Observational Sciences, Manora Peak, Nainital 263002, India.} 

\author{Sharmila Rani}
\author{Annapurni Subramaniam}
\affiliation{Indian Institute of Astrophysics, Koramangala, Bangalore 560034, India.}

\affiliation{Indian Institute of Astrophysics, Koramangala, Bangalore 560034, India.}

\collaboration{4}{}
\author{Gaurav Singh}
\affiliation{Indian Institute of Astrophysics, Koramangala, Bangalore 560034, India.}

\author{Snehalata Sahu}
\affiliation{University of Warwick, Coventry CV4 7AL, United Kingdom.}

\author{ Ravi S. Singh}
\affiliation{ Deen Dayal Upadhyay Gorakhpur University, Gorakhpur, Uttar Pradesh 273009,India.}

\email{arvind@aries.res.in}
\email{rkant@aries.res.in}
\email{purni@iiap.res.in}
\email{sharmila.rani@iiap.res.in}


\begin{abstract}
 
We report the discovery of extremely low-mass white dwarfs (ELM WDs) as a companion of blue straggler stars (BSSs) in the Galactic globular cluster NGC\,362 using images from \textit{AstroSat}'s Ultra Violet Imaging Telescope (UVIT). Spectral Energy Distributions (SEDs) for 26 FUV bright member BSSs are created using data from the UVIT, UVOT, \textit{Gaia} EDR3, and the 2.2 m ESO/MPI telescope. A single SED is fitted to 14 BSSs, whereas double-SED fits revealed ELM WDs  as binary companions in 12 of the 26 BSSs studied. The effective temperature, radius, luminosity and mass of the 12 ELM WDs are found to have a range (T$_{eff}$ $=$ 9750$-$18000 $K$, R $=$ 0.1$-$0.4 R$_{\odot}$, L $=$ 0.4$-$3.3 L$_{\odot}$, and M $=$ 0.16$-$0.20 M$_{\odot}$).  
These suggest that 12 BSSs are post-mass-transfer systems formed through Case A/B mass transfer pathway. To the best of our knowledge, this is the first finding of ELM WDs as companions to BSS in globular clusters. This cluster is known to have a binary BSS sequence and the 12 binary and 14 single BSSs (as classified by the SEDs) follow the mass transfer and collisional sequence of BSS in the colour-magnitude diagram. The cooling ages of 9 of the ELM WDs are found to be younger than 500 Myr. Though the binary BSSs may have formed during the core-collapse ($\sim$ 200 Myr) or as part of the dynamical evolution of the cluster, they provide new insights on the dynamics of this cluster. \\

\end{abstract}


\section{Introduction}
\label{sec:intro}
Globular clusters (GCs) are one of the oldest stellar systems in which frequent gravitational interactions among stars take place due to their high densities. 
These gravitational interactions lead to several dynamical processes, i.e., two-body relaxation, mass segregation from equipartition of energy  \citep{2007ApJ...663.1040L, 2014MNRAS.442.3105M, 2017MNRAS.464.2174B}, core-collapse \citep{1997A&ARv...8....1M}, stellar collisions, and mass transfer and mergers \citep{2022MNRAS.512.2936K} in binary star systems. These dynamical processes produce exotic populations like low-mass millisecond pulsars, cataclysmic variables, and blue straggler stars (BSSs). BSSs are found in bulk among these exotic populations  \citep{1995ApJ...439..705B, 2001ApJ...561..337F}.

BSSs were first discovered by \cite{1953AJ.....58...61S} in the outskirts of GC M\,3, where they appear as an extension of the main-sequence (MS) in the optical color-magnitude diagram (CMD). They are generally found in diverse stellar environments such as GCs \citep{1953AJ.....58...61S, 1993AJ....106.2324F}, open clusters (OCs) \citep{1955ApJ...121..616J}, dwarf galaxies \citep{2011ApJ...735...37C}, and  galactic fields \citep{2000AJ....120.1014P}. Several observational evidences \citep{1988ARA&A..26..199R, 1997ApJ...489L..59S} suggest that BSSs are more massive (M $\sim$ 1.2 M$_{\odot}$) than  the typical turn-off stars (M $\sim$ 0.95 M$_{\odot}$) in GCs. 

The formation of BSSs cannot be explained through canonical stellar evolution theory. Therefore, their formation must be looked into by some mechanisms that are able to increase the initial mass of single stars \citep{2018ApJ...860...36F}. The formation mechanism of BSS is still being debated. However, the two leading formation mechanisms of BSS are as follows: stellar collisions \citep{1976ApL....17...87H, 1989AJ.....98..217L} leading to a merger in high-density environments and mass transfer (MT) \citep{1964MNRAS.128..147M,1976ApJ...209..734Z} from an evolved donor to a lower-mass binary star in a low-density environment. Another mechanism for twin BSS formation in a compact binary system has been proposed by \cite{2019ApJ...876L..33P}, in which mass is transferred through a circumbinary disk from an evolved outer tertiary companion.

Depending on the evolutionary type of the donor star, three distinct scenarios of MT have been presented \citep{1967MComP...7..129K, 1971ARA&A...9..183P} for the formation of BSSs through Roche lobe overflow.
Case A: main sequence (MS) star, Case B: red giant branch (RGB) star, and Case C: asymptotic giant branch (AGB) star. In case A, a mass transfer can result in the production of either single BSS or short-period binary BSSs, whereas BSSs with a white dwarf (WD) companion having either He or CO core of masses 0.2-0.4 and 0.5-0.6 M$_{\odot}$, respectively are produced through Case B and Case C \citep{2009ApJ...697.1048P}.

Extremely low-mass (ELM ) helium white dwarfs  have effective temperatures in the 8000 K $\le$ T$_{eff}$ $\le$ 22000 K range and masses of less than 0.3 M$_{\odot}$. They are most likely the result of binary interactions \citep{1995/mnras/275.3.828}. These low-mass stellar objects cannot be formed from single-star progenitors because their nuclear evolution timescale exceeds the Hubble time unless they have extremely high metallicity \citep{2007ApJ...664.1088K}. Hence, the majority of these objects are found in binary systems with companions such as neutron stars in milli-second pulsar (MSP) systems \citep{2005ASPC..328..357V}, A/F-type dwarf stars \citep{2014MNRAS.437.1681M} or another (typically a carbon-oxygen) WD. These objects can be found in a variety of environments, including the Galactic disk, OCs and GCs \citep{2015/mnras/stv1810}. Hence, they can be formed from progenitors of varying metallicities. These WDs are believed to be originated through either unstable mass loss via common envelope (CE) events or stable mass loss via Roche-lobe overflow in close binary systems with a thick envelope \citep{2013A&A...557A..19A}.

In the CE model, the secondary component fills the Roche-lobe (RL)  during the Hertzsprung gap (HG) or near RGB phase leading to an unstable mass transfer. Such a state evolves into a new CE state, where the system eventually evolves into a double degenerate (DD) binary.
The ELM WDs with mass $\leq$ 0.2 M$_{\odot}$ cannot be produced through the CE scenario since the high binding energy of the CE in such a binary system could lead to the merger rather than the ejection during the CE evolution. In contrast, in the RL model, the secondary component initiates a stable MT during late MS, leading to the formation of a proto-He WD. The system then progresses to a DD binary with an ELM WD mass ranging from 0.14 to 0.30 M$_{\odot}$.

\cite{2013ApJ...778..135D} identified two distinct BSS sequences in the post-core-collapse GC NGC\,362. They suggested that this feature is caused by short-duration dynamical events in the cluster, such as core collapse. Similarly, the double sequences of BSS have been found in several GCs, including M\,30  \citep{2009Natur.462.1028F},  NGC\,1261 \citep{2014ApJ...795L..10S} and M\,15 \citep{2019ApJ...876...87B}. The bluer BSS sequence is thought to be formed by the increased star collision rate during core contraction, while the redder BSS sequence is caused by the increased Roche-lobe overflows.

UV observations are crucial to identify and understanding the nature of the companion to the BSSs. BSSs define a clear vertical sequence in the UV CMD \citep{2014AJ....148..131S, 2018ApJ...860...36F}. In particular, ASTROSAT/UVIT data has been extensively used to characterize BSSs and identify their hot companions based on UV excess. \cite{2016ApJ...833L..27S} discovered a post-AGB/HB companion of a BSS in the old OC NGC 188 utilizing UVIT data. \cite{2019ApJ...882...43S}, \cite{2021MNRAS.507.2373P} and \cite{2021JApA...42...89J}  used UVIT data to analyze OCs and found BSS with hot companions. In the GC NGC 1851, \cite{2020ApJ...905...44S}  discovered extreme Horizontal Branch (EHB) stars as hot companions to the BSS. 

This paper presents the UV photometric results of the GC NGC\,362 located in the southern hemisphere of the Tucana constellation and slightly northwards of the Small Magellanic Cloud (SMC).
The age of the cluster is $\sim$ 11 Gyr, located at a  distance of $\sim$ 8.83 kpc \citep{2021MNRAS.505.5978V},  with a reddening $\sim$ 0.05 mag \citep{2010arXiv1012.3224H} and a  metallicity of [Fe/H]  $\sim$ $-1.3$ dex.  The tidal radius (r$_{t}$) of the cluster is  $r \sim 16^{\prime}.11$.  This cluster has been extensively studied in the optical region, but UV study is limited. 
Our main aim in this study is to characterize the BSS population primarily using data from UVIT onboard \textit{AstroSat} to put more light on their nature as well as formation scenarios of BSSs in GC NGC\,362.

This paper is organized as follows: Section~\ref{sec:data} covers the archival/observational data along with their reduction and analysis methods. 
In section~\ref{sec:CMD}, we present the details of the UV and optical CMDs as well as the selection of BSSs. The details of the SED fitting technique are described in section~\ref{SED}. We discuss the properties of BSSs derived using the SED technique and their evolutionary status in section~\ref{sec:bss}. The detailed discussion of all the results is presented in section~\ref{sec:dis}, followed by the summary and conclusions in section~\ref{sec:conclusions}.

\begin{center}
\begin{table*}[]
    \caption{Log of archival data.}
    \begin{tabular} {p{3.20cm} p{6.6cm} p{2.35cm} p{4.0cm}  }
    \hline
    Telescopes  &	 Filters  &	FOV  &	Exposure Time(ksec)\\
    \hline
  HST  & F275W, F336W, F438W (WFC3/UVIS)& $160'' \times 160''$ & $3.1, 1.4, 0.2$  \\           & F606W, F814W(ACS/WFC) & $202'' \times 202''$ & $0.6, 0.6$\\
    UVIT/AstroSat & F148W, F169M, N245M, N263M&  $ 28' $ & $4.9,4.6,4.9, 4.6$\\
   UVOT/Swift    & uvw2(192nm), uvm2(224nm), uvw1(260nm)&$ 17' \times 17'$ & $2.4, 1.7, 1.7$ \\
    2.2m MPI/ESO      & U, B, V, R, I  & $ 34' \times 33' $ & $0.24, 0.24, 0.09, 0.045, 0.045$\\
    \hline\end{tabular}
    \label{obserbational table}
\end{table*}
\end{center}
\section{Data Sets and their reduction}
\label{sec:data}
In order to study the BSS population in NGC\,362, we have utilized the multi-wavelength data from several telescopes covering the entire cluster region up to r$_t$. We have used UVIT, UVOT, and 2.2 m MPI/ESO observations for this analysis. The details of the data used and their reductions are as follows:\\

a) UVIT : We used images of NGC\,362 captured by the UVIT instrument on board the \textit{AstroSat} satellite. UVIT is made up of two 38 cm telescopes, one for the FUV region (130-180 nm) and another for the NUV (200-300 nm) and visible (VIS) (320-550 nm) regions. The cluster's level-1 data was acquired from the \textit{AstroSat} archive \footnote{https://astrobrowse.issdc.gov.in/astro archive/archive/Home.jsp}. The observations of the NGC\,362 were carried out on 11$^{th}$ November 2016 in four UV filters: F148W, F169M, N245M, and N263M. The exposure time and filter information are provided in Table 1. Data reduction of the raw images was performed using CCDLAB (\cite{2017PASP..129k5002P}), which corrects the satellite drift, flat-field, distortion, fixed pattern noise, and cosmic rays. The detailed descriptions of the telescope, instruments and preliminary calibration can be found in \cite{2016SPIE.9905E..1FS} and \cite{2017AJ....154..128T}. 

The point spread function (PSF) photometry was performed on all UVIT images using DAOPHOT II \citep{1987PASP...99..191S} programs. The ALLSTAR program was then used to fit the PSF model to the aperture magnitudes of all known stars. The aperture correction was established in each filter using a curve-of-growth analysis and then applied to the estimated PSF magnitudes. The saturation correction \citep{2017AJ....154..128T} was applied to the PSF-generated magnitude to account for more than one photon per frame, yielding the instrumental magnitudes. Stars brighter than 17 mag are affected mainly by saturation correction. The zero-points are used from \cite{2017AJ....154..128T} to determine the magnitudes in the AB system. The astrometry of the UVIT images was performed using the WCS task in CCDLAB software \citep{2017PASP..129k5002P} by utilizing the sky coordinate information from {\it Gaia} eDR3.\\

b) UVOT: The raw  data obtained from the HEASARC archive\footnote{https://heasarc.gsfc.nasa.gov} have been processed using HEASoft\footnote{https://heasarc.gsfc.nasa.gov/docs/software/heasoft} pipeline. The log of observations is given in Table \ref{obserbational table}. The corrections were applied for exposure maps and auxiliary spacecraft data, and the science images were geometrically corrected for sky coordinates. The procedure described in \cite{2014AJ....148..131S} was followed to reduce the UVOT/Swift data. The large-scale sensitivity (LSS) maps for individual frames were generated using the UVOTSKYLSS task. These frames were combined into a single multi-extension file using the FAPPEND task. UVOTIMSUM generated a single image for each filter and their corresponding sensitivity and exposure map images. 

The PSF photometry on all UVOT images (uvw1, uvm2, and uvw2 filters) was performed using DAOPHOT II \citep{1987PASP...99..191S} programs. We performed aperture photometry with a $5^{\prime\prime}.0$ aperture radius to figure out the aperture correction. \cite{2020ApJ...905...44S} discussed the reduction and calibration of UVOT images in detail. \\

c) $2.2-$m ESO/MPI: The archival images obtained by the Wide-Field Imager (WFI) placed on the 2.2-m ESO/MPI ground-based telescope were used to investigate the outer region of NGC\,362. The broadband U, B, V, R, and I filters are used. The observational log of the archival data is provided in Table \ref{obserbational table}. With a total FOV of $34^{\prime}$ $\times$ $33^{\prime}$, the WFI is composed of eight CCDs (each with 2048 $\times$ 4096 pixels with a size of $0^{\prime\prime}$.238 pixel$^{-1}$). The raw science images were pre-processed using the MSCRED package available in IRAF.

The photometry on the WFI images was performed following the approach described in A06 \citep{2006A&A...454.1029A}. The PSF's shape varies greatly with position; hence, 9 PSFs per CCD chip ($3\times$3) were developed to limit spatial fluctuation to less than 1 \% \citep{10.1093/mnras/sty2961}. The PSFs were developed using an empirical grid with a quarter-pixel resolution. Each PSF was represented by a 201$\times$201 grid with a radius of 25 pixels and a center (101, 101). An automated code  \citep{2006A&A...454.1029A} was used to interactively find the instrumental magnitudes and positions using the array of generated PSFs. The code first detects the brightest stars and then moves to the fainter stars. The distortion solution derived in A06 was used to minimize the effect of geometric distortion. The B, V, R, and I instrumental magnitudes were converted to the standard magnitude using the Stetson \citep{2019yCat..74853042S} photometric standard. The image coordinates of WFI and UVOT images were converted into sky coordinates using the CCMAP and CCTRAN tasks of IRAF. The astrometric accuracy of all the images was $\sim$ $0^{\prime\prime}.1$. 
\begin{figure}
    \centering
    \includegraphics[width=\columnwidth]{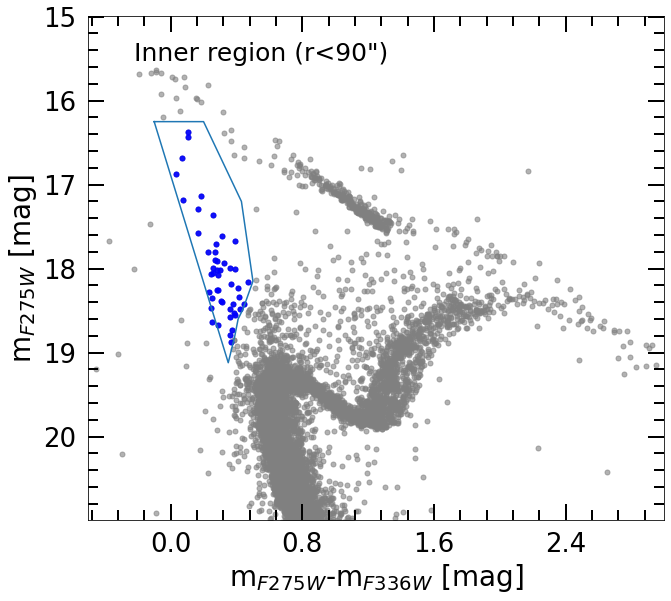}
    \caption{The selection criteria for the BSSs in UV-CMD (m$_{F275W}$, m$_{F275W}$-$m_{F336W}$) for the inner region is shown with box. The selected BSSs are shown with filled blue circles.}
    \label{figure:CMD_HST}
\end{figure}
\begin{figure}
    \centering
    \includegraphics[width=\columnwidth]{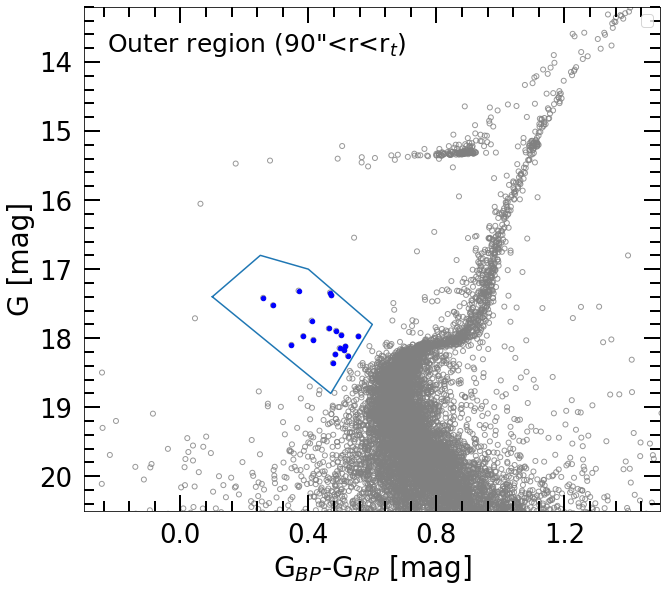}
    \caption{The selection criteria for the BSSs in Optical-CMD (G, G$_{BP}$ $-$ G$_{RP}$) for the outer region is shown in the box. The selected BSSs are shown with the filled blue circles.}
    \label{figure:CMD_GAIA}
\end{figure}
Apart from the above data, we also have used Hubble Space Telescope (HST) data covering the central region ($r \leq 90^{\prime\prime}$ ) of the cluster as it is an ideal instrument to resolve the core of the clusters. The \textit{HST} UV Legacy Survey of Galactic Globular Clusters (HUGS) catalog was used, which comprises photometric magnitudes in five filters, namely, F275W, F336W, F438W, F606W, and F814W along with proper-motion membership information \citep{2018MNRAS.481.3382N}.

\section{UV and Optical Colour Magnitude Diagrams} \label{sec:CMD}
In this section, we present the details of data cross-match and selection of BSSs using optical and UV CMDs. The UVIT data were combined with UV-optical astro-photometric data from the  HST, UVOT, 2.2m ESO/MPI, and \textit{Gaia} EDR3. The matched UV-optical data were analyzed using the UV-optical CMDs for selecting BSSs.
\begin{figure}
	\includegraphics[width=\columnwidth]{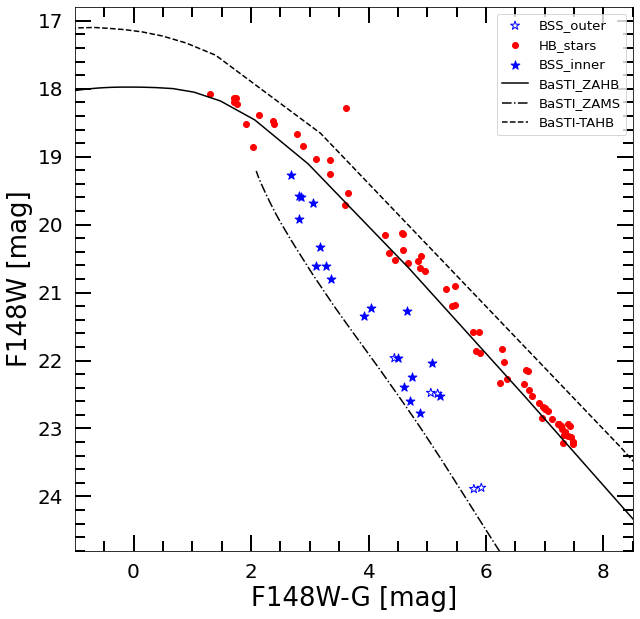}
    \caption{The F148W vs F148W$-$G CMD of hot populations. The UVIT-Gaia EDR3 cross-matched sources are plotted. The reddening and distance modulus of the cluster are taken from \cite{2010arXiv1012.3224H}. The models of ZAMS, ZAHB and TAHB taken from BaSTI-IAC online database \citep{2018ApJ...856..125H, 2021ApJ...908..102P} are over-plotted.}
    \label{FUV_G}
\end{figure}
\begin{figure*}
    \centering
    \includegraphics[width=180mm,height=9.5cm]{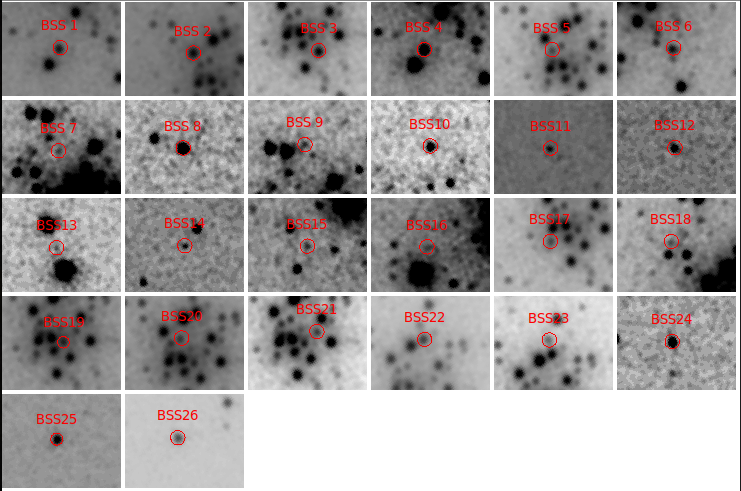}
    \caption{The location of FUV bright BSS in FUV (F148W) image of UVIT. Each image has a field of view of $40'' \times 30''$.}    
    \label{figure:image}
\end{figure*}
\subsection{Cross-match of UV data with optical data} \label{archive}
Our main focus in this paper is to study the UV-bright BSSs detected in the FUV images. For this purpose, we first identified the FUV-bright BSSs in the inner region ($r\leq90^{\prime\prime}$) by cross-matching UVIT and HST photometric data with matching radius $1^{\prime\prime}$ . For identifying the FUV bright BSSs in the outer region (90$^{\prime\prime}<r<r_t$), we cross-matched the UVIT data with the UVOT/Swift and 2.2m ESO/MPI photometric data with matching radius $1^{\prime\prime}.5$. We used the membership probability information provided in the GlobULeS I catalog \citep{2022MNRAS.514.1122S} for the outer region. Stars having a membership probability greater than 90\% were chosen to be cluster members and considered for further analysis.

\subsection{Selection of Blue Straggler Stars }
Using optical counterparts of the UVIT detected sources in NGC\,362, we constructed UV-optical CMDs of members for the inner and outer regions. The CMDs are a vital tool to  identify stars at various stages of evolution in clusters. Both UV and optical CMDs can be used to select the BSSs in star clusters, but UV CMDs are preferred over optical CMDs as BSSs appear brighter in them and also can be easily separated from the hot HB stars  \citep{2017ApJ...839...64R} and stars near the MS turn off.

In Figure~\ref{figure:CMD_HST}, we plotted  NUV CMD (m$_{F275W}$, m$_{F275W}$ $-$ m$_{F336W}$) using the HST catalog. For the outer region, we created the optical CMD (G, G$_{BP}$ $-$ G$_{RP}$) using {\it Gaia} eDR3 photometric catalog as shown in Figure~\ref{figure:CMD_GAIA}. In both CMDs, dots represent stars with a membership probability greater than 90\%. 

The BSSs were chosen using the method described in \cite{2017ApJ...839...64R}, with UV-CMD provided as the main selection criterion. To select the BSS in both the inner and outer regions of the cluster, we drew a box around the BSS sequence in the CMDs presented in Figure~\ref{figure:CMD_HST} and Figure~\ref{figure:CMD_GAIA}. We defined the fainter limit of the box in Figure~\ref{figure:CMD_HST} as 5 $\sigma$ above the MS turn-off. By averaging the magnitudes and colors of stars in a box of 1 mag near the turn off point on the MS, $\sigma$ was calculated. In this way, a total of 48 BSSs were identified in the inner region. A similar box was also drawn in Figure~\ref{figure:CMD_GAIA} to pick out the BSSs in the outer region. We found 20 BSSs in the outer region. In total, 68 BSSs (48 in the inner and 20 in the outer region) were identified in the cluster NGC\,362 and are depicted with blue filled circles in Figure~\ref{figure:CMD_HST} and Figure~\ref{figure:CMD_GAIA}.

To clearly distinguish and select the FUV bright BSSs, we plotted the FUV-Optical CMD (F148W, F148W$-$G) shown in Figure~\ref{FUV_G}. The red-filled circles represent the HB stars, while the filled and open blue symbols represent the BSSs for the inner and outer regions, respectively. In this way, we selected 26 FUV bright BSSs. We also plotted the updated BaSTI-isochrones \footnote{http://basti-iac.oa-abruzzo.inaf.it} \citep{2018ApJ...856..125H, 2021ApJ...908..102P} models of zero-age main sequence (ZAMS), zero-age HB (ZAHB), and terminal-age (TAHB) in the CMD with dash-dot, solid, and dashed lines respectively. These isochrones are corrected for reddening and extinction \citep{1999PASP..111...63F, 2019A&A...632A.105C}. Both HB and BSS sequences can be clearly seen in Figure~\ref{FUV_G}. The BSS sequence is $\sim$ 4 mag while the HB sequence is $\sim$ 5 mag stretched along the F148W magnitude axis. All these FUV bright BSS are clearly resolved, as shown in Figure~\ref{figure:image}. \\

\section{Spectral Energy Distribution of the FUV bright BSS} 
\label{SED}
\begin{figure*}[]
\setkeys{Gin}{width=\linewidth}
\includegraphics{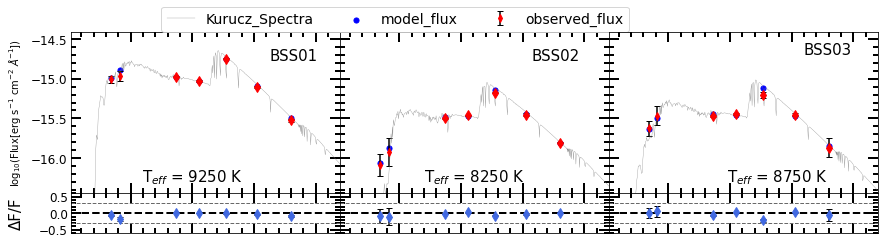}
\includegraphics{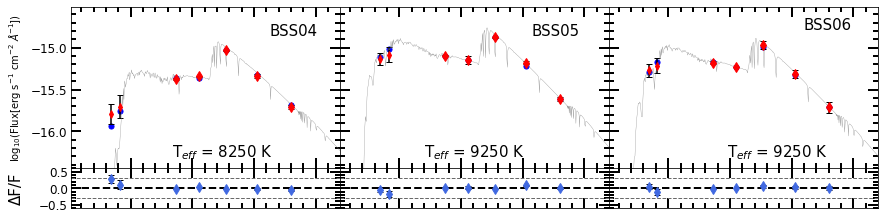}
\includegraphics{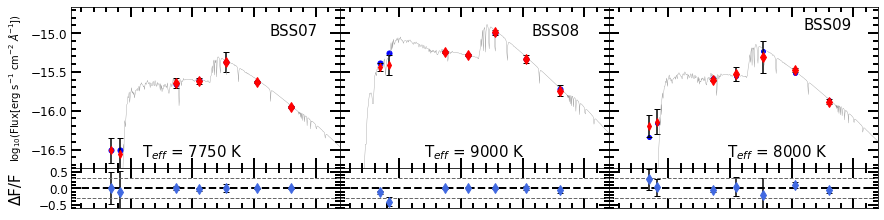}
\includegraphics{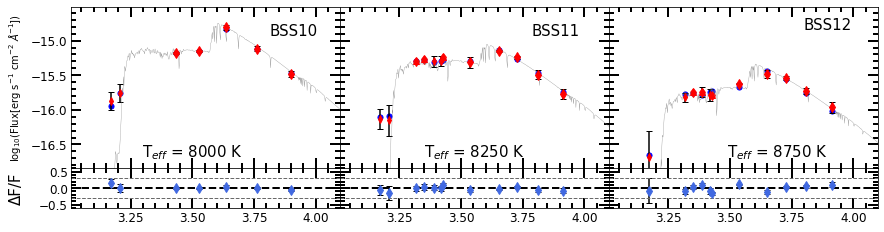}
\includegraphics{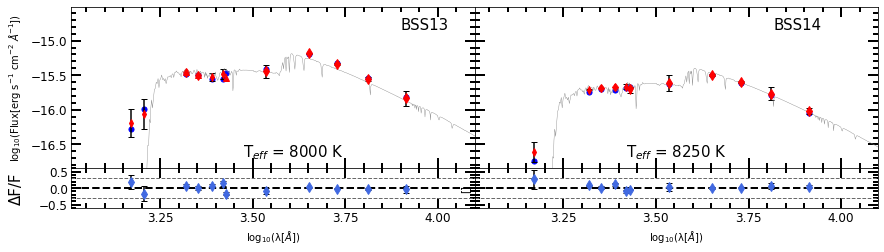}

\caption{The SEDs of single-component BSSs. The blue filled circles represent the synthetic flux from the model used to fit the observed SED of single-component BSSs. The best-fit atmospheric parameters are mentioned in Table \ref{table:single_fit_para}. The residuals of the SED fit are presented in the bottom panel of all plots.}
\label{SED_single}

\end{figure*}

\begin{table*}[]
    \centering
    \caption{The best-fit parameters of single-component BSSs. The positions (RA, DEC) are given in degrees. The T$_{eff}$ is the effective temperature in K, log $g$ is the surface gravity in logarithm unit, the reduced chi-square ($\chi^2_{r}$), the luminosity (L/L$_{\odot}$) and radius (R/R$_{\odot}$) is in Solar unit, Vgfb and Vgfb are the visual goodness of fit, N$_{fit}$ is the number of points considered in the fitting and N$_{tot}$ is the total number of points.}
    \begin{tabular}{p{1.3cm} p{1.2cm} p{1.35cm} p{1.5cm} p{0.8cm} p{0.8cm} p{1.8cm} p{1.8cm} p{1cm} p{0.5cm} p{1cm} }
    \hline
     Name & RA & DEC &	T$_{eff}$ (K) &	Log $g$  &	$\chi^2_{r}$ &	L/L$_{\odot}$ & R/R$_{\odot}$ &	Vgf &	Vgfb & N$_{fit}$/N$_{tot}$ \\
    \hline
  BSS01& 15.81374&	$-$70.85611  & 9250$^{+500}_{-250}$ & 4.5 & 9.92 & 19.140$^{+0.170}_{-0.320}$ & 1.718$^{+0.100}_{-0.180}$ & 22.10 & 0.99 & 7/7\\
  BSS02& 15.81938&	$-$70.84622 & 8250$^{+250}_{-750}$ & 4.0 & 10.60 & 7.424$^{+0.001}_{-0.299}$ & 1.343$^{+0.220}_{-0.073}$ & 5.65 & 0.66 & 7/7\\
  BSS03&  15.80834&	$-$70.85098	& 8750$^{+250}_{-1000}$& 4.0 & 7.24 & 7.537$^{+0.136}_{-0.219}$ & 1.210$^{+0.292}_{-0.104}$ & 23.60 & 1.48 & 7/7\\
  BSS04& 15.79833&	$-$70.85612 & 8250$^{+250}_{-750}$ & 4.0 & 11.70 & 9.847$^{+0.313}_{-0.247}$ & 1.541$^{+0.368}_{-0.203}$ & 15.70 & 2.39 & 7/7\\
  BSS05& 15.81351&	$-$70.84771	& 9250$^{+500}_{-500}$ & 4.5 & 13.70 & 14.860$^{+0.300}_{-0.030}$ & 1.493$^{+0.181}_{-0.133}$ & 15.70 & 1.03 & 7/7\\
  BSS06& 15.79856&	$-$70.85527 & 9250$^{+250}_{-500}$ & 5.0 & 11.03 & 11.690$^{+0.680}_{-0.030}$ & 1.325$^{+0.255}_{-0.065}$ & 6.99 & 0.47 & 7/7\\
  BSS07& 15.81503&	$-$70.84177 & 7750$^{+500}_{-250}$ & 4.5 &  4.92 & 4.970$^{+0.079}_{-0.030}$ & 1.236$^{+0.070}_{-0.121}$ & 3.73 & 3.01 & 7/7\\
  BSS08& 15.83265&	$-$70.83091 & 9000$^{+250}_{-500}$ & 4.5 & 27.82 & 10.740$^{+0.370}_{-0.320}$ & 1.356$^{+0.239}_{-0.118}$ & 23.30 &  4.49 & 7/7\\
  BSS09& 15.81316&	$-$70.83779 & 8000$^{+250}_{-500}$ & 4.0 & 10.22 & 6.324$^{+0.110}_{-0.358}$ & 1.311$^{+0.199}_{-0.163}$ & 10.20 &  2.54 & 7/7\\
  BSS10& 15.7891&	    $-$70.83161& 8000$^{+250}_{-500}$ & 4.0 & 21.82 & 15.920$^{+0.160}_{-1.000}$ & 2.074$^{+0.296}_{-0.322}$ &  15.00 &  1.81 & 7/7\\ 
  BSS11& 15.82898&	$-$70.81353 & 8000$^{+250}_{-500}$ & 5.0 & 27.24 & 8.920$^{+0.349}_{-0.307}$ & 1.568$^{+0.362}_{-0.100}$ & 18.40 & 3.99 & 11/12\\
  BSS12& 15.63451&	$-$70.89542 & 8250$^{+250}_{-500}$ & 4.5 & 27.46 & 7.756$^{+0.539}_{-0.13}$ & 1.371$^{+0.367}_{-0.091}$ &  17.30 & 3.45 & 11/12\\
  BSS13& 15.65826&	$-$70.81532	& 7500$^{+250}_{-750}$ & 4.0 & 23.94 & 4.538$^{+0.375}_{-0.097}$ & 1.181$^{+0.331}_{-0.092}$ & 18.50 & 1.53 & 11/12\\
  BSS14& 16.07722&	$-$70.92675 & 7750$^{+750}_{-250}$ & 5.0 & 26.69 & 4.369$^{+0.124}_{-0.273}$ & 1.151$^{+0.107}_{-0.228}$ & 13.10 & 2.16 & 11/11\\
    \hline\end{tabular}
    \label{table:single_fit_para}
\end{table*}
\begin{figure*}[h]
\setkeys{Gin}{width=\linewidth}
\includegraphics{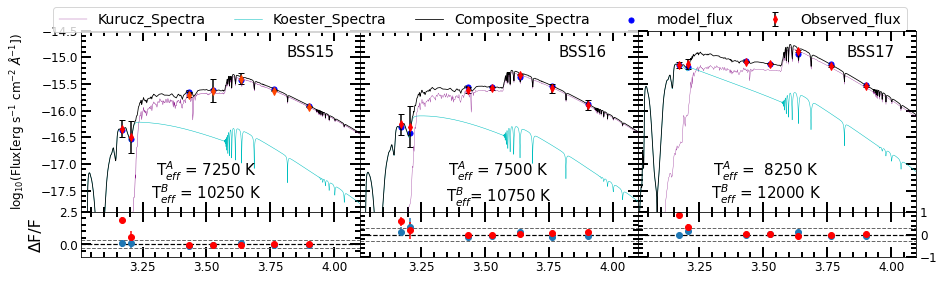}

\includegraphics{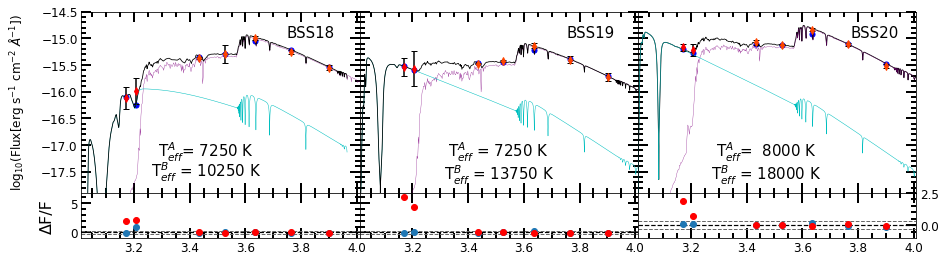}

\includegraphics{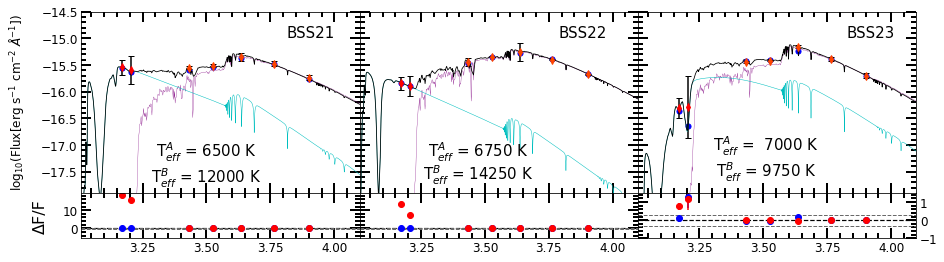}

\includegraphics{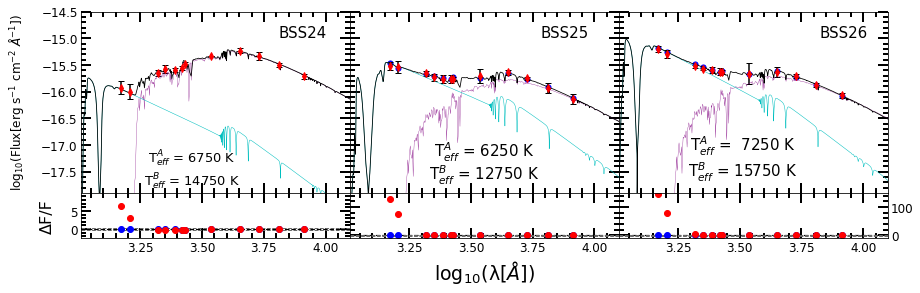}

\caption{The SEDs of double-component BSSs using the data taken from UVIT, UVOT, and 2.2m ESO/MPI telescopes after correcting for extinction. The composite SED fit to the BSSs is shown with black colour while purple and cyan color spectra represent the Kurucz and Koester model as displayed in the legend, respectively. The best-fit parameters are displayed in Table \ref{table:binary_fit_para}.}
\label{SED_binary}
\end{figure*}

\begin{table*}[h]
    \centering
    \caption{The best-fit parameters for the double-component BSSs. Headers are the same as described in Table \ref{table:single_fit_para}. Here, ``A" denote the cool component (BSS), and ``B" denote the hot component (ELM WD).}
    \begin{tabular}{p{1.3cm} p{1.2cm} p{1.35cm} p{1.5cm} p{1cm} p{0.8cm} p{1.7cm} p{1.7cm} p{1cm} p{0.6cm} p{0.9cm} }
    \hline
     Name & RA & DEC &  	T$_{eff}$ (K) &	         Log $g$  & $\chi^2_{r}$ & L/L$_{\odot}$ &           R/R$_{\odot}$ &	Vgf &	Vgfb & N$_{fit}$/N$_{tot}$ \\ 
    \hline
  BSS15 A & 15.84851 &	$-$70.85769 &  7250$^{+125}_{-125}$   & 4.00  & 6.059 & 4.543$^{+0.045}_{-0.040}$  & 1.353$^{+0.010}_{-0.010}$  & 5.86 & 4.73 & 7/7\\
  BSS15 B &          &              & 10250$^{+750}_{-250}$  & 6.50  & 6.059 & 0.393$^{+0.078}_{-0.065}$  & 0.199$^{+0.029}_{-0.033}$  & 5.86 & 4.73 & 7/7\\
  BSS16 A & 15.83432 &  $-$70.85095 &  7500$^{+125 }_{-125}$  & 4.00 & 1.415 & 5.075$^{+0.120}_{-0.164}$   & 1.335$^{+0.013}_{-0.024}$  & 1.25 & 0.91 & 7/7\\
  BSS16 B &          &              & 10750$^{+500 }_{-250}$ & 7.25 &  1.415 & 0.498$^{+0.157}_{-0.098}$  & 0.204$^{+0.076}_{-0.036}$  & 1.25 & 0.91 & 7/7\\
  BSS17 A & 15.81471 &  $-$70.85065 &  8250$^{+125 }_{-125}$  & 4.00  & 27.300 & 12.790$^{+0.080}_{-0.770}$  & 1.753$^{+0.005}_{-0.041}$  & 12.40 & 0.95 & 7/7\\
  BSS17 B &          &              & 12000$^{+250 }_{-1000}$& 7.25 &  27.300 & 3.286$^{+0.538}_{-0.095}$  & 0.420$^{+0.125}_{-0.023}$  & 12.40 & 0.95 & 7/7\\
  BSS18 A & 15.82657 &  $-$70.84255 &  7250$^{+125 }_{-125}$  & 3.50  & 3.600 & 10.800$^{+0.160}_{-0.190}$  & 2.086$^{+0.009}_{-0.136}$  & 2.57 & 0.96 & 7/7\\
  BSS18 B &          &              & 10250$^{+500 }_{-250}$ & 6.50  & 3.600 & 0.877$^{+0.121}_{-0.220}$  & 0.298$^{+0.034}_{-0.057}$  & 2.57 & 0.96 & 7/7\\
  BSS19 A & 15.80813 &  $-$70.84887	&  7250$^{+125 }_{-125}$  & 3.50  & 13.760 & 07.646$^{+0.000}_{-0.900}$  & 1.766$^{+0.210}_{-0.008}$  & 9.17 & 1.75&7/7\\
  BSS19 B &          &              & 13750$^{+750 }_{-250}$ & 6.50  & 13.760 & 1.241$^{+0.069}_{-0.050}$  & 0.195$^{+0.009}_{-0.023}$  & 9.17 & 1.75 & 7/7\\
  BSS20 A & 15.81073 &  $-$70.84724 & 8000$^{+125 }_{-125}$  & 4.00  & 8.532 & 14.910$^{+0.220}_{-0.430}$  & 2.040$^{+0.007}_{-0.035}$  & 8.53 & 2.11 & 7/7\\
  BSS20 B &          &              & 18000$^{+750 }_{-750}$& 6.50  & 8.532 & 2.149$^{+0.321}_{-0.207}$  & 0.151$^{+0.085}_{-0.018}$  & 8.53 & 2.11 & 7/7\\
  BSS21 A & 15.80635 &	$-$70.84908 & 6500$^{+125 }_{-125}$  & 3.50  & 6.984 & 5.995$^{+0.116}_{-0.151}$  & 1.931$^{+0.019}_{-0.014}$  & 4.85 & 1.12 &7/7\\
  BSS21 B &          &              & 12000$^{+500 }_{-250}$ & 7.00  & 6.984 & 1.722$^{+0.214}_{-0.001}$  & 0.302$^{+0.006}_{-0.032}$  & 4.85 & 1.12 & 7/7\\
  BSS22 A & 15.80396 &	$-$70.84564 & 6750$^{+125 }_{-125}$  & 5.00  & 3.186 & 8.032$^{+0.190}_{-0.010}$  & 2.046$^{+0.009}_{-0.003}$  & 3.02 & 1.17 & 7/7\\
  BSS22 B &          &              & 14250$^{+750 }_{-250}$ & 6.50  & 3.186 & 0.614$^{+0.076}_{-0.001}$  & 0.127$^{+0.001}_{-0.019}$  & 3.02 & 1.17 & 7/7\\
  BSS23 A & 15.80928 &	$-$70.8421  & 7000$^{+125 }_{-125}$  & 3.50  & 9.313 & 7.256$^{+0.175}_{-0.199}$  & 1.833$^{+0.030}_{-0.097}$  & 5.94 & 2.75 &7/7\\
  BSS23 B &          &              & 9750$^{+500 }_{-250}$  & 6.75  & 9.313 &  1.187$^{+0.160}_{-0.109}$ & 0.382$^{+0.041}_{-0.082}$  & 5.94 & 2.75 &7/7\\
  
  BSS24 A & 15.67558 &	$-$70.78341 & 6750$^{+125 }_{-125}$  & 5.00  & 21.15 & 08.157$^{+0.100}_{-1.608}$    &2.112$^{+0.004}_{-0.003}$ & 11.5 & 0.57 & 12/12\\
  BSS24 B &          &              &14500$^{+250 }_{-1000}$& 6.50  & 21.15 & 00.455$^{+0.014}_{-0.015}$    &0.107$^{+0.012}_{-0.001}$ & 11.5 & 0.57 & 12/12\\
  BSS25 A &15.74852  &	$-$70.89148 & 6250$^{+125 }_{-125}$  & 5.00  & 09.35 & 02.331$^{+0.038}_{-0.101}$    &1.308$^{+0.012}_{-0.029}$ & 8.02 & 0.87 & 12/12\\
  BSS25 B &   &                     &12750$^{+750 }_{-250}$ & 7.50  & 09.35 & 01.376$^{+0.026}_{-0.043}$    &0.241$^{+0.028}_{-0.012}$ & 8.02 & 0.87 & 12/12\\
  BSS26 A & 15.69907 & $-$70.82305 & 7250$^{+125 }_{-125}$  & 5.00  & 26.45 & 02.824$^{+0.085}_{-0.055}$    &1.064$^{+0.003}_{-0.011}$ & 17.1 & 1.12 & 12/12\\
  BSS26 B &   &                     & 15750$^{+250 }_{-500}$ & 7.75  & 26.45 & 02.312$^{+0.010}_{-0.132}$    &0.206$^{+0.016}_{-0.093}$ & 17.1 & 1.12 & 12/12\\

\hline\end{tabular}
\label{table:binary_fit_para}
\end{table*}

A detailed SED analysis was performed to evaluate the fundamental parameters of the FUV bright BSSs, such as their total luminosity (L/L$_{\odot}$), effective temperature (T$_{eff}$), and radius (R/R$_{\odot}$). The SED was created using photometric data ranging from UV to IR wavelengths for the 26 FUV bright BSS candidates. SED analysis was carried out using the Virtual Observatory SED Analyzer (VOSA; \cite{2019hsax.conf..430R}). 

To correct for extinction in the observed data points, VOSA employs the reddening relations provided by \cite{1999PASP..111...63F} and \cite{2005ApJ...619..931I}. The extinction-corrected VOSA magnitudes agree with those calculated by us. The  VOSA performs a reduced chi-square ($\chi^2_{r}$) minimization test by comparing synthetic photometry to observed data to determine the best-fit SED parameters. \cite{2016ApJ...833L..27S} and \cite{2021MNRAS.501.2140R} describe the detailed procedure for analyzing SEDs. VOSA provides two more characteristic parameters, $Vgf$ and $Vgf_{b}$, in addition to $\chi^2_{r}$, which are referred to as visual goodness of fit. $Vgf$ and $Vgf_{b}$ should be less than 25 and  15 respectively, \citep{2018MNRAS.480.4505J, 2021MNRAS.506.5201R} for the best fit. VOSA estimates the uncertainties in the derived parameters using Markov Chain Monte Carlo (MCMC) technique.

The Kurucz stellar atmospheric models were used to generate synthetic SEDs for the FUV bright BSSs \citep{1997A&A...318..841C, 2003IAUS..210P.A20C}. The effective temperatures were chosen to be between 5000 – 50000 K, with a metallicity value of $[Fe/H]$ = $-1$ close to the cluster metallicity, and a corresponding log $g$ ranging from 3$-$5 dex. For the SED analysis of FUV BSSs, the  UVIT data were combined with HST data for the inner region and with UVOT and {\it Gaia} data for the outer region.

First, we fitted single component SED to the 26 FUV bright BSSs with Kurucz stellar atmospheric model as shown in the top panel in Figure~\ref{SED_single} and \ref{SED_binary}. The data points used in the fitting are shown as red, cyan, and orange-red-filled circles with error bars, while model data points are shown as blue points. The Kurucz spectra are depicted with a grey line. The residual between the fitted model and the observed fluxes normalized by the observed flux across all filters is shown in the bottom panel of each SED. The residual points with the error are represented by sky blue-filled circles. The dashed horizontal lines drawn at $\pm$ 0.3 (30\%) represent the residual's threshold. The residual plots presented in Figure~\ref{SED_single} show that the residuals for the 14 BSSs (BSS01 - BSS14) are within 0.3 across all wavelengths. This indicates that the single-component Kurucz model fits these 14 BSSs well.

To determine the fundamental parameters of the 14 BSSs (Group 1), VOSA uses $\chi^2_{r}$ fitting. Table \ref{table:single_fit_para} shows the effective temperature, log $g$, luminosity, and radius along with the $\chi^2_{r}$ for the single-component BSSs. The range in different parameters are found as  T$_{eff}$ $\sim$ 7500 - 9250 K, log $g$ $\sim$ 4.0 - 5.0, L$\sim$ 4.3 - 19.1 L$_{\odot}$ and R $\sim$ 1.1 - 2.0 R$_{\odot}$. We can see that estimated parameters are consistent with the BSS properties \citep{2016ApJ...833L..27S, 2021MNRAS.501.2140R}. 

Figure~\ref{SED_binary} shows the best SED fits for the 12 BSSs (BSS15 - BSS26). The Kurucz model is  shown with a gray solid line. The residual plot for the 12 BSSs (Group 2) shows that the residual flux in the UV region is greater than 0.3 in more than one data point. This indicates that there is UV excess, and the SEDs may be fitted by a combination of hotter and cooler spectra.

\subsection{The Two-component SED fits} 
\label{Binary_sed}
To account for the UV excess in the BSS of Group 2, we fitted two-component SEDs as shown in Figure~\ref{SED_binary}. We used the Kurucz stellar atmospheric model \citep{1997A&A...318..841C, 2003IAUS..210P.A20C} to fit for the cooler component and the Koester WD model \citep{2010MmSAI..81..921K} to the hotter component. In the Koester WD model, the T$_{eff}$ and log $g$ range from 5000$-$80000 K and 6.5$-$9.5, respectively. 

In the top panel, observed data points are indicated with red filled circles, while synthetic data points are shown with blue-filled circles. The cool and hot synthetic spectra are represented by grey and black lines, respectively. We used VOSA to get the two-component fit of these stars \footnote{http://svo2.cab.inta-csic.es/theory/vosa50/helpw4.php?otype
=star\&action=help\&what=fitbin}. The composite SEDs is shown with black color in Figure \ref{SED_binary}. We can see that the composite SEDs are well-fitted across all wavelengths. The hot components are well-fitted with a Koester model, whereas the cool components are well-fitted with a Kurucz model. In the bottom panel of each SED, the fractional residuals are plotted with sky blue points. The fractional residuals are within $\pm$0.3 (30\%) across all wavelengths. 

The fundamental parameters of these BSSs derived using composite SED fits are listed in Table \ref{table:binary_fit_para}. The range in parameters for cool companions are T$_{eff}$ $\sim$ 6200 - 8250 K, log $g$ $\sim$ 3.5 – 5.0, L$\sim$ 2.3 – 14.9 L$_{\odot}$, and R $\sim$ 1.0 – 2.1 R$_{\odot}$, and for hot companions, T$_{eff}$ $\sim$ 9750 - 18000 K, log $g$ =6.5 - 7.75, L $\sim$ 0.4 - 3.3 L$_{\odot}$ and R $\sim$ 0.1 - 0.4 R$_{\odot}$. The goodness of fit parameters, i.e., $\chi^2_{r}$, $Vgf$, and $Vgf_{b}$ are also listed in Table \ref{table:binary_fit_para}. Based on their $T_{eff}$ and radius, we believe that the cool companions can be BSSs, and the hot companions can be WDs. 


\section{Blue straggler stars}
\label{sec:bss}
\subsection{ Mass and Age Estimation of BSSs} \label{mass_bss}
\begin{figure}
	\includegraphics[width=\columnwidth]{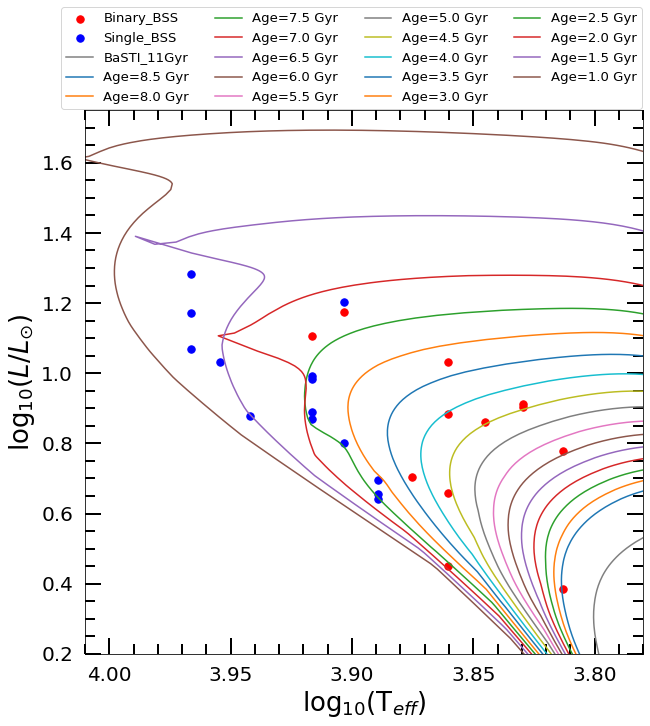}
    \caption{ The HR diagram ( L (L$_{\odot}$), T$_{eff}(K)$ ) for 26 BSSs. The Group 1 and 2 BSSs are represented  with blue and red colors, respectively. The theoretical isochrones taken from BaSTI \citep{2018ApJ...856..125H, 2021ApJ...908..102P} for [$\alpha/Fe]=0.4$ are represented by the continuous lines with a step of 0.5 Gyr age.}
    \label{Fig:BSS_mass}
\end{figure}

The mass of the BSSs is critical for understanding their formation mechanisms. We constructed the H-R diagram of BSS, which is shown in Figure~\ref{Fig:BSS_mass}. The Group 1 (blue) and Group 2 (red) BSSs are shown as filled circles. To determine the mass of BSS, we have plotted BaSTI evolutionary tracks \citep{2018ApJ...856..125H, 2021ApJ...908..102P} depicted with solid lines in the HR diagram as shown in Figure~\ref{Fig:BSS_mass}.
The theoretical evolutionary tracks have ages ranging from 1 to 8.5 Gyr with 0.5 Gyr steps. The location of BSSs in the CMD suggests that they are more massive than typical GC stars. Their masses are calculated by comparing their positions with the overlaid tracks. The derived masses and ages of 26 BSSs are $\sim$ 1.0 - 1.6 M$_{\odot}$ and 1.5 - 8.5 Gyr, respectively. 

We also note that the blue (Group 1) and the red (Group 2) BSS are found to occupy separate areas in the HR Diagram and could make two sequences. This is discussed below.

\begin{center}
\begin{table}[]
    \caption{The mass and cooling age (in Myr) for the ELMs.}
    \begin{tabular}{lll}
    \hline
   Name & Mass (M$_{\odot}$) & Cooling Age \\ 
    \hline
  BSS15 B    & 0.164-0.170          &  1285-1400\\
  BSS16 B    & 0.170-0.176          &    480-580 \\
  BSS18 B  & 0.176-0.182           &  1285-1400\\
  BSS23 B  & 0.182-0.186            &  755-875 \\
  BSS22 B, BSS24 B &0.186-0.192        & 460-475\\
  BSS19 B, BSS21 B, BSS25 B & 0.192-0.202  & 300-500 \\
  BSS17 B, BSS20 B, BSS26 B &  0.202         & 200-400\\
    \hline
    \end{tabular}
    \label{table:ELM}
\end{table}
\end{center}

    \begin{figure}
	\includegraphics[width=\columnwidth]{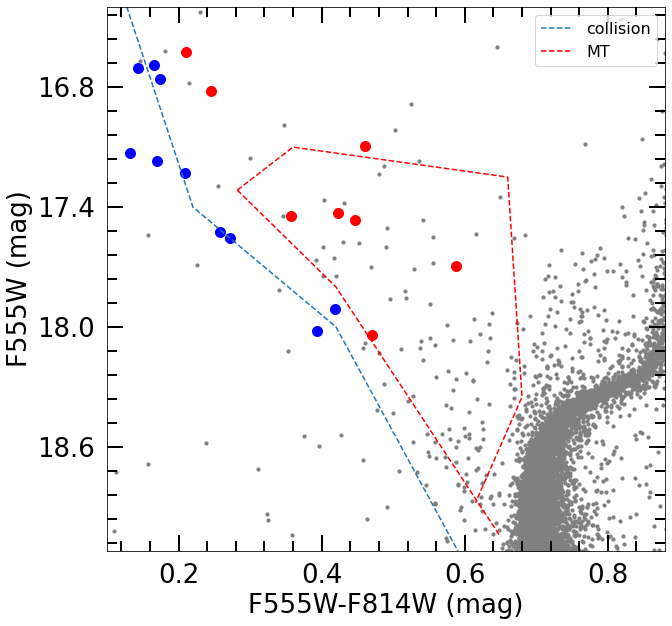}
    \caption{The F555W, (F555W $-$ F814W) CMD for the FUV bright BSSs. The Group 1 and 2 BSSs are plotted with red and blue points. The dotted blue line is a 0.2 Gyr collisional isochrone, and the red dotted box indicates the region of MT binaries taken from D13.}
    \label{Fig: V_vs_I_reproduce.png}
    \end{figure}

    \begin{figure}
	\includegraphics[width=\columnwidth]{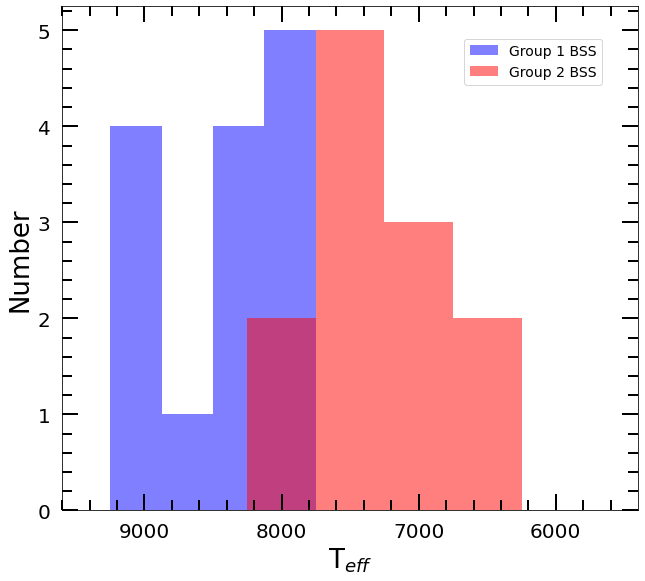}
            \caption{The number distribution of BSSs with effective temperature. The Group 1 and 2 BSSs are highlighted with blue and red colors, respectively.}
    \label{Fig:Hist}
     \end{figure}

\subsection{BSS Double Sequence } \label{Double_seq}

\cite{2013ApJ...778..135D} (D13) discovered a double BSS sequence in NGC\,362. On the basis of a comparison of the observed BSS distribution with theoretical models, they have suggested the presence of two parallel BSS sequences. They claim that the blue sequence is primarily formed by collisions, whereas the red sequence is formed through a continuous MT formation process. Based on our SED analysis, we classify Group 1 as single and Group 2 as binary BSS.
To compare the position of Group 1 and Group 2 BSS with the CMD of D13, we converted F606W ACS magnitude to F555W WFC3 magnitude using the equations given in \cite{2017ApJ...835...28J}. It is obvious that our Group 1 and Group 2 BSS locations in the CMD correspond to the blue and red groups of D13.

In Figure~\ref{Fig: V_vs_I_reproduce.png}, we plotted Group 1 and 2 BSS with blue and red filled circles, respectively. The blue curve represents a collisional isochrone of 0.2 Gyr age \citep{2009ApJ...692.1411S}, while the orange curve represents the mass transfer model. The blue line and orange box correspond to the same region defined by D13 for collisional and  MT BSSs, respectively. This diagram shows that Group 1 BSSs adhere to the collisional model, whereas Group 2 BSSs occupy the region defined for the BSSs formed via mass transfer. This is consistent with the conclusions made in D13. Therefore, we can say that Group 1 BSS formed through collision, whereas Group 2 BSSs formed through mass transfer. We found 5 FUV bright BSS brighter than 17 magnitude in this analysis, as shown in Figure ~\ref{Fig: V_vs_I_reproduce.png}. D13 did not identify these BSS in their study. Figure ~\ref{Fig: V_vs_I_reproduce.png} also suggests that these bright BSSs are following a blue and red sequence.

To check the temperature distribution of Group 1 and Group 2 BSS, we created a histogram of effective temperature for all BSSs, as shown in Figure~\ref{Fig:Hist}. Group 1 and 2 BSS are represented by blue and red colours, respectively. Group 1 BSS has a significantly higher temperature than Group 2. This indicates that collisional BSSs have higher temperatures as compared to MT BSSs. We did not perform the radial distribution of Group 1 and Group 2 BSS because the number of FUV bright BSS was insufficient.

\begin{figure}
	\includegraphics[width=\columnwidth]{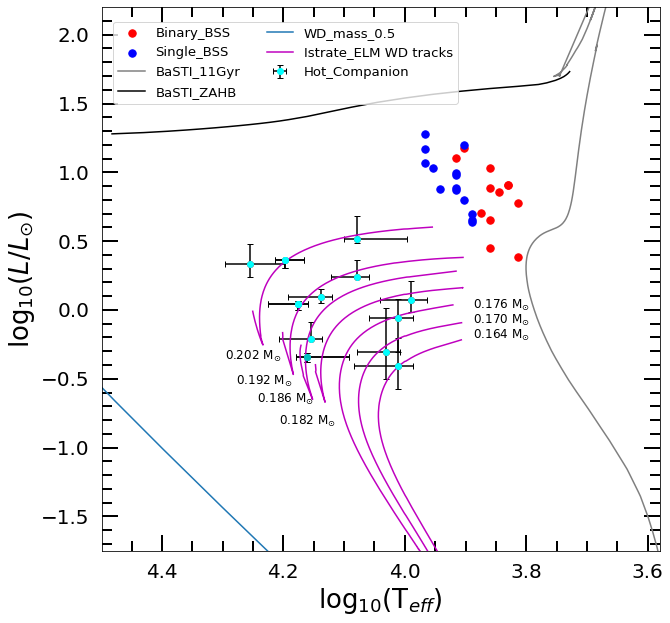}     
    \caption{The H-R diagram with both BSS (single component BSS, the cooler component of the two-component BSS ) represented as blue and red filled circles respectively, and the hotter components of Group 2  denoted by cyan circles. The BaSTI isochrones and BaSTI ZAHB  is depicted as  a brown and black curve with the cluster's age  11 Gyr. The ELM WD curves \citep{2016A&A...595A..35I} are plotted as magenta solid color lines with masses 0.151 – 0.202 M$_{\odot}$.}
    \label{Fig: HR}
\end{figure}

\subsection{H-R diagram} \label{HR_diagram}

We already discussed the locations of the BSSs in the H-R diagram in section~\ref{mass_bss}. To study their properties, we over-plotted the BaSTI isochrones and BaSTI ZAHB for the 11 Gyr age with black lines. The BSSs are clearly visible in this diagram at the extension part of MS and beneath the ZAHB. The H-R diagram of hot companions of the Group 2 BSSs is shown in Figure~\ref{Fig: HR}. Along with the error bars, the hot companions are represented by cyan-filled circles. The blue line represents the WD cooling curve for mass 0.5 M$_{\odot}$. This diagram shows that the hot companions are located between MS and WD cooling curves and are much less luminous than the HB stars. Based on the location of the hot companions in the HR diagram, they could be ELM WDs. To determine the mass and cooling age of ELM WDs, we over-plotted the ELM WD cooling curves \citep{2016A&A...595A..35I} with magenta color for the mass range of 0.165-0.202 M$_{\odot}$. Using the closest ELM WD cooling curve, the cooling age and mass of individual ELM WDs can be estimated, and the same is listed in Table \ref{table:ELM}. Table \ref{table:ELM} infers that the cooling age and mass range of ELM WDs are  200 - 1400 Myr and 0.164 - 0.202 M$_{\odot}$, respectively. \\

\section{Discussion} 
\label{sec:dis}
We have identified 26 FUV bright BSSs in NGC\,362 based on the data taken from UVIT/AstroSat, HST, and {\it Gaia} eDR3. The BSSs are selected up to r$_t$ of the cluster. In our analysis, 14 BSSs are found as single stars, and 12 BSSs have companions as ELM white dwarfs. The BSSs and ELM white dwarfs are characterized using the SEDs. Discussions on the findings are stated below.\\

     The SED analysis of 26 BSSs was carried out by fitting the Kuruz stellar atmospheric model. SEDs analysis revealed that 14 BSSs are single stars.  We can deduce that they formed through a collisional process by comparing them to the model. However, three BSS are located in the cluster's outskirts. The origin of the outer BSS is unknown. \cite{1993ApJ...408L..89B}  and  \cite{1994ApJ...431L.115S} have suggested that the outer BSS may have been formed by collisions during resonance encounters between binary and single stars in the core, and ejected into the outer regions by the recoil from the interactions. On the other hand,  \cite{1995ApJ...439..705B} suggested that the outer BSSs have formed from mergers of primordial binaries, similar to the BSS in sparse clusters. To know a clear picture of the formation of outer BSS, both of these explanations require elaboration. Both mechanisms appear to play an important role in the formation of outer BSSs in NGC\,362.
    
     In section \ref{mass_bss}, we estimated the mass of each BSS. The heaviest mass  BSS fitted through single SED is 1.64 M$_{\odot}$.  Furthermore, by comparing with the model in section \ref{Double_seq}, we demonstrated that these BSSs formed via the collisional process. Assuming that the initial mass of the two BSS progenitor stars is the same as the cluster's turnoff mass (0.8 M$_{\odot}$), these two stars may collide and merge to form the cluster's heaviest mass BSS.

     12 out of 26 BSSs were identified as binary systems using the SEDs. The cool and hot components of the binary system were characterized by Kurucz and Koester model. The physical properties derived in this analysis for the hot components indicate that they could be ELM WDs. We note that the Group-1 BSSs are found to have ages of $\sim$ 1.5, $\sim$ 2.5 Gyr as seen from Figure~\ref{Fig:BSS_mass}. The BSS+ELM pairs are younger than the younger sub-group of the Group-1 BSSs.
    
     We discovered that  BSSs fitted with a double component model have ELM WDs as companions in this study. The mass of a WD produced by a single star is limited to 0.4 M$_{\odot}$ \citep{2010ApJ...723.1072B}. The age of the universe is also a limiting factor at the lower end of the WD mass. ELMs, on the other hand, can be found in binary form in the universe (\cite{2019ApJ...883...51R}, \cite{2019ApJ...886...13J}, \cite{2020JApA...41...45S}, \cite{2022MNRAS.511.2274V}). The lower limit on the mass of the ELM WD can be attributed to mass loss during the early stages of evolution. The Case A/B MT may result in the He-core WD via early envelope mass loss \citep{1991ApJS...76...55I,1995/mnras/275.3.828}. Therefore, MT is essential for the creation of ELM WDs in tight binary systems where the companion tears away the ELM WD progenitor's envelope and the low mass core fails to ignite the He core. The companion star gained mass and converted to BSS during the mass loss.
    
    The mass range of detected ELM WDs is 0.16 - 0.20 M$_{\odot}$. According to the \cite{2021A&A...646A..30A}, the mass threshold for the occurrence of CNO flashes on the cooling branch of ELM WDs is 0.2 M$_{\odot}$. Therefore, no ELM WDs detected in the current study should produce CNO flash on the cooling branch. According to \cite{2019ApJ...871..148L}, ELM WDs with masses less than 0.22 M$_{\odot}$ formed from the Roche lobe overflow channel. This further concludes that these BSSs are formed via MT with ELM WD as their companion.
    
    In the present analysis, the maximum and minimum masses of the BSSs formed through the MT channel are 1.40 and 1.02 M$_{\odot}$, respectively. By assuming that the initial mass of the BSS progenitor is the same as the cluster's turn-off mass (0.8 M$_{\odot}$), the highest and lowest masses acquired by BSS from the binary system's primary are 0.6 and 0.2 M$_{\odot}$, respectively. The lower limit of mass is possible to be  acquired through an efficient Case-B MT, whereas the higher limit demands the progenitor of the ELM to be more massive than the turn-off mass so that more mass is available for transfer, or the present BSS went through more than one event of mass acquisition to gain up to 0.6 M$_{\odot}$.   
    \\
       
       The implication of finding the BSS+ELM WD pairs in a GC would mean that these binary systems remained dynamically stable to complete the MT process. The cooling age of the ELM WDs suggests that they can be as young as 200 Myr to as old as 1.4 Gyr, where 9 of them are as younger than $\sim$ 500 Myr. This cluster is understood to have undergone core-collapse 200 Myr ago \citep{2022arXiv221102671C}.  If we assume that this cluster is a post-core-collapse cluster, then it is possible that the cluster experienced an increase in the  number of binaries as a result of the core collapse. We speculate that these systems may be the result of the evolution of such binaries, as the cooling ages of the ELM WD companions and the time of core-collapse are not very different.
       
        \cite{2022MNRAS.511.4060K} found that in all of their simulations, the overall WD binary fraction, as well as the WD–MS binary fraction increases over a time-scale of 10 Gyr. They also find that a good number of Helium WDs are produced in their simulations, that can be formed only in binaries. They speculate that more frequent dynamical interactions force binaries to form Helium WDs because of mass transfer. The number of BSSs also increases with time, in their simulations. These ELM WDs, which are indeed helium WDs, as companions to BSSs are the first observational detections of such systems found in simulations. It is quite possible that we have detected probably the brightest of the ELM WDs and many more may be present.
       
       Therefore, the post-mass transfer binary systems (BSS+ELM WDs) could be formed either due to core-collapse or due to the overall gradual dynamical evolution of the cluster. Either way, the detections of the BSSs with ELM WD companions found in this core-collapsed cluster have implications to constrain the dynamical evolution theories of GCs.
       
       We do not attempt to study the radial distribution of these systems as our data is incomplete in the central regions. In order to increase the sample of FUV-detected BSS, we plan to carry out deeper observation of this cluster using UVIT for possible detection of fainter WD companions to BSSs.

\section{Summary and Conclusions}
\label{sec:conclusions}
In this work, we present the first detection of 12 BSSs with ELM WDs companions in the GC NGC\,362. The analysis is based on the UVIT observations as well as archival data from UVOT, {\it Gaia} eDR3 and the 2.2m ESO/MPI telescope. 
 The following are the main findings of this study.
\begin{enumerate}

    \item We have identified 26 FUV bright BSSs, which are confirmed PM members, utilizing the UVIT and {\it Gaia} eDR3 data. 
    
    \item The stellar parameters of the FUV bright BSSs were derived using SED fitting technique. Out of 26 BSSs, 14 have been successfully fitted with a single temperature SED. The effective temperature, radius, and luminosity of the 14 BSSs range from 7500$-$9250 K, 1.1$-$2.0 R$_{\odot}$, and 4.3$–$19.1 L$_{\odot}$, respectively.

    \item 12 BSSs were found to exhibit UV excess. The UV excess was more than 30\% compared to the model. Therefore, a double component SED was fitted to these BSSs and a range in stellar parameters (T$_{eff}$ $=$ 9750$-$18000 K, R $=$ 0.1$-$0.4 R$_{\odot}$, and L $=$ 0.4$–$3.3 L$_{\odot}$) for the hot components (ELM WDs) were estimated. The cool component's (BSSs) temperature, radius, and luminosity were estimated to be T$_{eff}$ $=$ 6250$-$8250 K, R $=$ 1.0$-$2.1 R$_{\odot}$, and L $=$ 2.3$–$14.9 L$_{\odot}$.  

    \item The masses of the hot components were found to be 0.16 $-$ 0.20 M$_{\odot}$. We classify the hot components as ELM WDs based on their location in the HR diagram. For the first time in a GC, we detect ELM WDs as companions to BSSs formed via the Case A/B mass transfer pathway.
    
    \item The mass of the BSSs was determined in the range of 1.0 $-$ 1.6 M$_{\odot}$ using the evolutionary tracks. We identified the double BSS sequence in the H-R diagram based on the SED analysis. In comparison to the model, our estimates support that the hotter BSSs are formed via a collisional process, whereas the cooler BSSs are formed via a mass transfer pathway. \\
    
    \item The cooling ages of 9 of the ELM WD companions were estimated to be younger than 500 Myr, whereas this cluster is expected to have undergone core-collapse $\sim$ 200 Myr ago. These binary BSSs may be formed during the core-collapse or as part of the dynamical evolution of the cluster. The detections of these post-mass-transfer systems in this core-collapsed cluster, therefore, provide new constraints on its internal dynamics.

\end{enumerate}


\section{ACKNOWLEDGEMENTS}
I would like to thank Dr. Sindhu Pandey for her guidance in the data reduction of UVIT.  I also would like to express my thanks to Vikrant Jadhav, Deepthi S. Prabhu, Gurpeet Singh, and Namita Uppal for their help throughout the entire process. AS acknowledges support from the SERB Power Fellowship.

\bibliography{paper}{}
\bibliographystyle{aasjournal}

\end{document}